\documentclass[structabstract]{aa}

\usepackage{graphicx}
\usepackage{natbib}
\begin{document}

   \title{New deep coronal spectra from the 2017 total solar eclipse}
   
 \titlerunning{Deep solar corona spectra}


   \author{S. Koutchmy\inst{1} \and F. Baudin\inst{2} \and Sh. Abdi\inst{2,3} \and L. Golub\inst{4} \and F. S{\`e}vre\inst{1}}

   \institute{
        Institut d'Astrophysique de Paris, CNRS/UPMC, Paris, France
        \and Institut d'Astrophysique Spatiale, CNRS/Universit\'e Paris-Sud, Orsay, France
        \and Observatoire de Paris, Universit\'e PSL, Meudon, France
        \and Smithsonian Astrophysical Observatory, Cambridge MA, USA
             }

   \date{Received  / accepted }

 
  \abstract
 {The origin of the high temperature of the solar corona,  in both the inner bright parts and the more outer parts showing flows toward the solar wind, is not understood well yet. Total eclipses permit a deep analysis of both the inner and the outer parts of the corona using the continuum white-light (W-L) radiations from electrons (K-corona), the superposed spectrum of forbidden emission lines from ions (E-corona), and the dust component with F-lines (F-corona).}
   {By sufficiently dispersing the W-L spectrum, the Fraunhofer (F) spectrum of the dust component of the corona appears and the continuum Thomson radiation can be evaluated. The superposed emission lines of ions with different degrees of ionization are studied to allow the measurement of temperatures, non-thermal velocities, Doppler shifts, and abundances to constrain the proposed heating mechanisms and understand the origin of flows that lead to solar wind.}
   {We describe a slit spectroscopic experiment of high spectral resolution to provide an analysis of the most typical parts of the quasi-minimum type corona observed during the total solar eclipse of Aug 21, 2017 from Idaho, USA. Streamers, active region enhancements, and polar coronal holes (CHs) are measured well using deep spectra.}
  {Sixty spectra are obtained during the totality with a long slit, covering $\pm$3 solar radii in the range of 510\,nm to 590\,nm.  The K+F continuum corona is exposed well up to two solar radii. The F-corona can be measured even at the solar limb. New weak emission lines were discovered or confirmed. The rarely observed Ar\,X line is detected almost everywhere; the Fe\,XIV and Ni\,XIII lines are clearly detected everywhere. For the first time hot lines are also measured inside the CH regions. The radial variations of the non-thermal turbulent velocities of the lines do not show a great departure from the average values. No significantly large Doppler shifts are seen anywhere in the inner or the middle corona. The wings of the  Fe\,XIV line show some non-Gaussianity.}
   {Deep slit coronal spectra offered an opportunity for diagnosing several aspects of coronal physics during a well observed total eclipse without extended investments. The analysis of the ionic emission line profiles offers several powerful diagnostics of the coronal dynamics; the precise measurement of the F-continuum component provides insight into the ubiquitous dust corona at the solar limb.}


   \maketitle
%

 \section{Introduction}
 \label{sec:intro}

Total solar eclipses (TSEs) are rare events that permit the analysis of the solar corona with great contrast and a high signal to noise ratio (S/N) for both imaging and spectra due to the low level of scattered light. Even rarer is the occurrence of a TSE at an easily accessible site with a very clear sky throughout the event, thus permitting an in-depth analysis. 
 Some of the most significant historical discoveries in both chromospheric and coronal physics are the result of TSE observations \citep{Allen46,vdHulst53,Shklovskii65,Guillermier91,Pasachoff09,Landi16}. These include the following: evidence of high temperatures that often reach 2\,MK at a distance of typically 1\,arcmin above the limb; the abundance ratio of elements of different first ionization potentials (FIPs) with high degrees of ionization of forbidden emission lines in the corona \citep{Edlen43,Edlen69}; the flows suggested by the occurrence of extended streamers on W-L images; and Doppler shifts of its emission lines \citep{Kim00,Mierla08}. The peculiar turbulent profiles of coronal ionic emission lines suggest a possible heating mechanism along the magnetically dominated structures as confirmed later from space-borne off-disk observations \citep[e.g.,][]{Doyle98} and from ground-based coronagraphic observations \citep{Koutchmy83,Contesse04}. The existence of a dusty component superposed along the line of sight (l.o.s) established \citep{Allen46} down to the limb of the Sun \citep{Koutchmy73}. 
 The mechanism at the origin of the sharp increase of the temperature immediately above the surface is still unknown, although the role of the magnetic field generated by dynamo effects in the deep layers is suggested to be a dominant actor in the higher layers where magneto-hydro-dynamics (MHD) processes take place. Several mechanisms of heating are actively debated, and many fundamental applications in stellar physics and even in several cosmological objects are of great interest. Briefly they can be classified as follows: first, turbulent dissipative processes related to ubiquitous magnetic reconnection events, sometimes called ``nano-flares'' \citep[e.g.,][]{Shibata11,Klimchuk15}; second, nonlinear dissipative processes \citep[for example the thermalization of slow mode magneto-acoustic waves magnified by the fast transverse Alfven and/or kink propagating coronal waves; shocks at interface between different layers with flows, etc., e.g.,][]{Osterbrock61,Cirtain13}; and third, the Joule effect in ubiquitous electric currents that are inferred from assumed circuits in loops \citep[e.g.,][]{Heyvaerts84,Ionson84}. 
Diagnostics permitting the measurement of very small scale events producing temperature effects on the emission measure (EM) variations and/or producing displacements (proper motions) and true velocities (Doppler shifts) are critically needed. Analyzing the forbidden coronal emission line profiles that reflect first, the ionic temperature, second, the non-thermal broadenings due to unresolved velocities, and third, the Doppler effect of unresolved structures, is an important diagnostic. Space-borne extreme ultra-violet (EUV) imagers provide excellent coronal images \citep[e.g.,][]{Parenti17}, but the continuum emission is not available. Deep EUV spectrograms that would show the K-continuum to measure the electron corona are not available. The same difficulty exists when ground-based Lyot coronagraphs are used \citep{Singh04} because of the large amount of scattered light. The K-continuum is then barely measurable \citep{Contesse04}; its polarization should potentially be used \citep{Landi16}.

The absence of a solid diagnostic, which is used to compute the plasma densities in a straightforward manner, is a serious deficiency of modern X-EUV space-borne observations. However, the spatial resolution reached is excellent today compared to the classical observations by OSO-7, for example \citep{Allen75}, when the summed structures measured along the l.o.s. were described using a filling factor. Narrow pass-band, high spatial resolution filtergrams are now supplied to complement spectra, and proxies are introduced based on the evaluation of the emission measures of permitted EUV emission lines provided the spectral resolution is sufficient for avoiding the overlapping of lines. Continuous X-ray and EUV observations at a high temporal rate are routinely available permitting a deep insight into the mechanism of heating. Regarding the flows in the more outer corona, externally occulted coronagraphs collect much lower resolution images taken in W-L at an excellent rate but unfortunately, the external occulter does not show the intermediate corona where flows presumably start.\\
At the rare TSEs where the continuum dominates in each deep spectrum, we have the opportunity to also measure the F-component superposed over the K-corona component where F-lines are almost completely washed out. This F-component, which is not to be confused with the scattered light from the terrestrial atmosphere \citep{Stellmacher74},  is due to the dust surrounding the Sun and originates in heliospheric comets heated by the Sun, degrading asteroids, and meteorites and small micron size bodies orbiting close to the Sun. At ground-based facilities large aperture Lyot coronagraphs that work under the direct solar light are limited by the parasitic light of different origins, and solar F-lines are imprinted \citep[e.g.,][]{Koutchmy83,Contesse04}. Their spatial resolution is often too marginal for a deep spectroscopic analysis of barely resolved structures due to turbulent effects of the daytime seeing and inside the instrument where the solar disk is concentrated on the artificial Moon. In space, a similar difficulty occurs due to scattered light of instrumental origin \citep{Mierla08}. The F-component, which is superposed on the K-continuum, cannot be clearly detected from the interplanetary dust.

During a TSE, the use of a small aperture amateur size telescope equipped with a good spectrograph and a modern camera performs order of magnitudes better for a deep analysis of coronal emissions. It offers a great opportunity, without a great investment, to deduce some fundamental parameters of coronal physics: an improvement of at least three orders of magnitudes is immediately available due to the efficient shadowing by the Moon of both the terrestrial atmosphere and the instrument during totality. Much deeper spectra can then be obtained \citep{Allen46,Nikolsky71,Jefferies71,Magnant73,Koutchmy73,Raju93,Ichimoto96,Mouradian97}. Modern CCD detectors go even further in permitting a very precise measurement of the background variations, as was done for the first time by \citet{Ichimoto96}. Unfortunately some parasitic light of terrestrial atmospheric origin limited the quality of their results. 

For the 2017 TSE, we prepared a spectroscopic experiment based on several attempts made with the classical fast emulsions since the 1973 African TSE when the photographic film method was still used \citep{Stellmacher74,Koutchmy74}; at the 1994 TSE observed from a high altitude site in Chile \citep{Koutchmy95}; and at the 1999 TSE observed from Iran and the TSE of 2001 observed from Angola with the first results obtained in the green Fe XIV line profiles up to more than 1\,R$_{\odot}$ from the limb \citep{Bocchialini01,Koutchmy05}. Some more recent attempts were made with the linear CMOS detector in 2010-2012 in French Polynesia (Hao) and in Australia (near Cairns), and finally in 2016 in Indonesia, using a computer-controlled CMOS detector camera, which had mixed success because of a thin layer of rather significant cirrus clouds that produced a lot of scattered light over the entire spectra due to the very bright fringe of the inner corona.\ These observing conditions were similar to those from the \citet{Ichimoto96} experiment. However, we could extract several high quality spectra\footnote{\begin{flushleft}Some qualitative spectra from the 1973, the 1994, and the 2016 TSE can be downloaded at https://www.facebook.com/photo.php?fbid=10214134899263842\& set=a.4438317433429\&type=3\&theater\end{flushleft}} as far as the spectral resolution is concerned \citep{Bazin13}.

\section{Observations}

\begin{figure*}
    \centering
    \includegraphics[height=7cm,width=18cm]{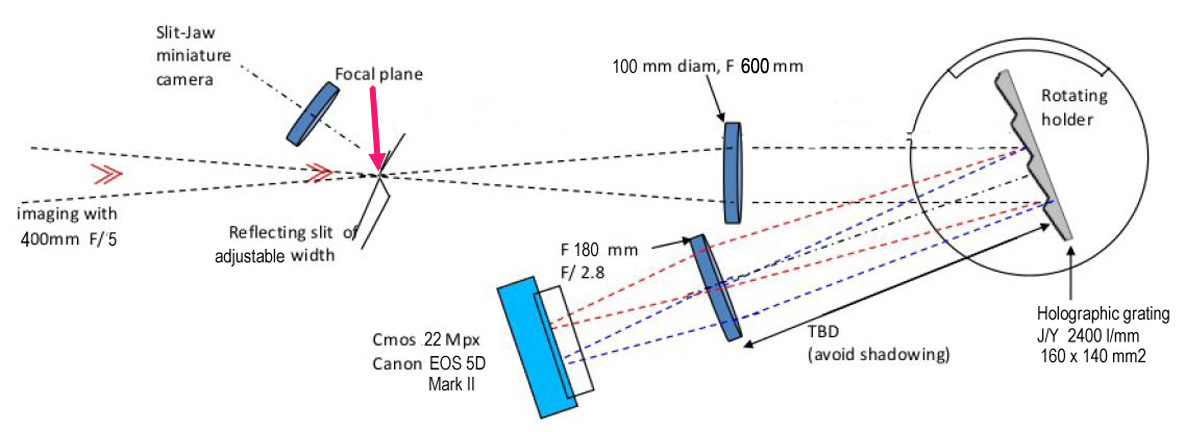}
    \caption{Schematic view of set-up used. The imaging doublet lens at left is not shown.}
    \label{fig:schema_spectro}
\end{figure*}

\begin{figure*}
    \centering
    \includegraphics[height=9cm,width=18cm]{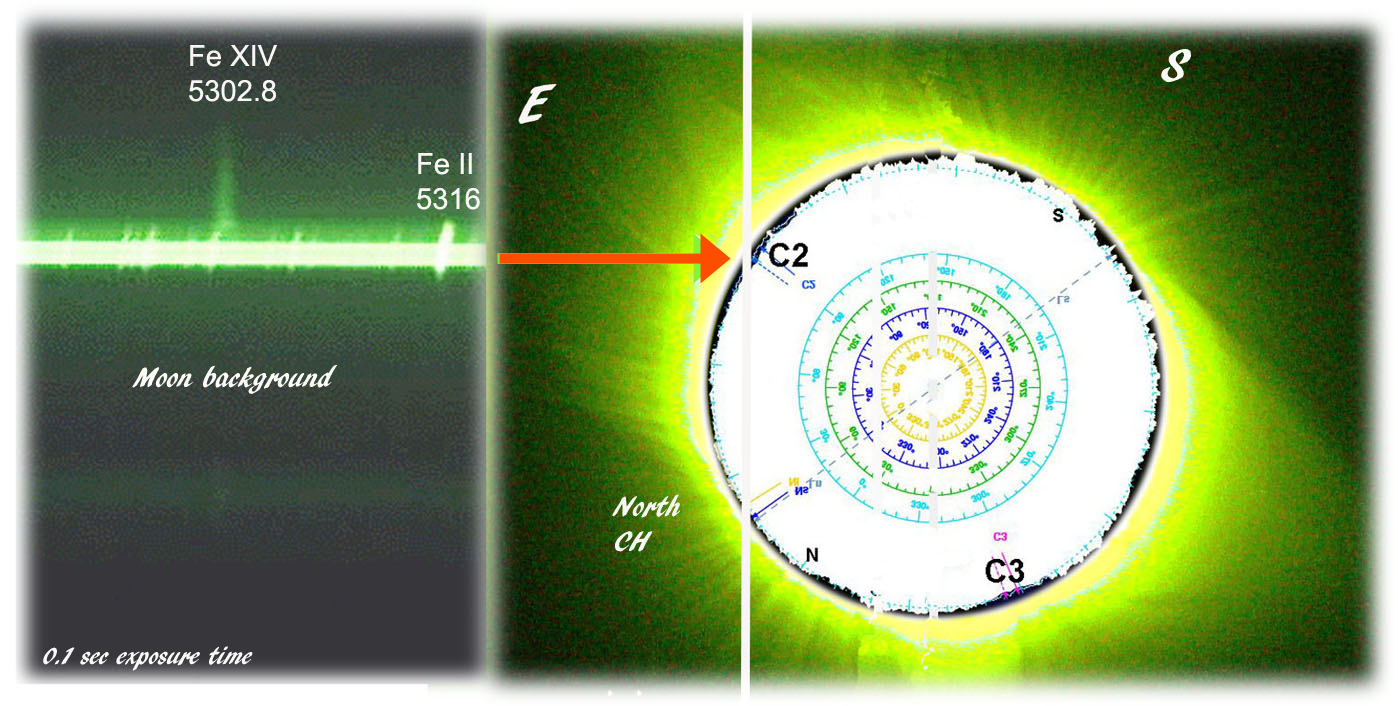}
    \caption{On left, image of selected part (1/35 of spectrum is shown) of single short exposure spectrum taken during C2 contact, with vertical slit position shown at right over unprocessed W-L image taken at same time with green filter; Jubier's diagram of edge of Moon is overlaid (see http://xjubier.free.fr/en/site\_pages/SolarEclipseCalc\_Diagram.html); amplitudes of limb lunar features are amplified by factor 80 and width of slit is not to scale. The C2 full spectrum shows hundreds of  faint low excitation emission lines formed in the minimum temperature region \citep[see][]{Bazin13} often with a low FIP line, such as the bright Fe\,II line shown on the right of the partial spectrum. In the blue wing of the Fe\,XIV coronal line (on the left), rising above the local edge of the Moon, the weak Fe\,I solar line is seen in emission instead of being in absorption, confirming the quality of our spectra. The F-component, which is fainter than the bright K-corona in the near equatorial regions, is not shown.}
    \label{fig:overview}
\end{figure*}

 For the 2017 TSE, the coordinates of the site of observation situated in Indian Valley (Idaho, USA) are: latitude 44$^{\circ}$\,26.47833\,N and longitude 116$^{\circ}$\,28.03167, which are at a distance of less than 20\,km from the central line of eclipse totality and an altitude of almost 1000\,m. The maximum of totality occurred at 17h\,27min\,09s\,UT (10h\,27min local time). The expected duration of totality was 2min\,6s and the height of the Sun was 45$^{\circ}$5'. The ratio of angular sizes Moon/Sun was 1.02774. The position of the Sun at the time of these observations is declination +11$^{\circ}$\,52'\,23'' and right ascension 10h\,03min\,56s with a bright star near the TSE ($\alpha$ Leo, or Regulus, of magnitude m = 1.5).\\
As noted above, the 2016 spectra confirmed the quality of our 600\,mm collimator focal length spectrograph with a 105\,$\mu$m wide entrance slit that uses a computer controlled Mark\,II Canon 5D camera to record the spectra (see Fig.\,\ref{fig:schema_spectro}). Raw \textit{.CR2} files are recorded with a header giving the parameters of each image, including the time of the exposure. We note that 5616\,$\times$\,3744\,px size images made of elementary 2\,$\times$\,2 Bayer matrices are recorded with a 14 bits precision using the 24\,$\times$\,36\,mm$^2$ CMOS chip. The elementary px size is 6.4\,$\mu$m and we used the sensitivity level of 800 ISO, giving a read-out noise of 8.27 ADU or 4.2 e$^-$. Both the dark noise level, the bias, and hot pixels are corrected as usual. For calibration purposes, solar spectra in diffuse W-L were taken using a white diffuser put before the entrance aperture before and after the totality with different levels of occulted Sun. This allowed us to measure the spectral response of the whole system and to get a reference F-spectrum. This spectral response is dominated by the behavior of the response of the camera (see Fig.\,\ref{fig:sensi}), and to a lesser extent, the grating efficiency (blaze). The response depends on the processing applied to the recorded elementary-pixels image that is called demosaicing or debayering. We applied our own program to deduce the full resolution spectra by only averaging in the direction of the length of the slit where the image is over-sampled in the spatial direction. The exit of the spectrograph uses a lens with a f/2.8 focus collimator after the dispersion of the light from our holographic 140$\times$160\,mm$^2$ Jobin-Yvon grating of 2400\,l/mm working in the first order (see Fig.\,\ref{fig:schema_spectro}). The grating is at the heart of our experiment; it is known that the ultimate flux of photons that reach each pixel of our camera critically depends on the effective size of the grating, and the achieved spectral resolution ultimately depends on the number of lines per mm and the size of the grating. The resulting dispersion at the exit of the spectrograph was on the order of 50\,$\mu$m/0.1\,nm due to the exit collimator of 180\,mm focus at f/2.8. The measured equivalent instrumental width ($\Gamma_{inst}$) that uses the narrow photospheric lines, while avoiding the blending of lines, of the solar spectra were taken in diffuse light just after the totality was 0.072\,nm. This is close to the optimal value when getting the higher possible S/N over the whole profiles of coronal emission lines without significantly affecting the spectral resolution.  Calibration spectra were also taken without a diffuser in front of the entrance aperture by using the thin crescent of the partially occulted Sun attenuated with an astro-solar filter of neutral density d= 3.7.\\
The spectrograph was pointed visually, watching the image of the corona reflected on the entrance slit of 105$\,\mu$m width to select the analyzed regions. Fig.\,\ref{fig:overview} is a composite to qualitatively illustrate the typical performance of the experiment. An image of a very small portion (3\% of a typical full spectrum) of a single short exposure frame obtained during the second contact (C2) is shown in the region of the Fe\,XIV line near 530.3\,nm. Fourteen frames were taken during C2 at a 3 frames/sec cadence using a 1/10\,s exposure time to analyze the chromosphere.

The width of the slit is comparatively large because of the priority given to the much weaker and broader coronal emission lines at large distances from the limb where the lines and the background due to the K and F corona are weak. Thus, we used a spectral width that is comparable to the typical FWHM of the bright green coronal line in order to achieve the optimum S/N. 

\begin{figure*}
    \centering
    \includegraphics[width=17.75cm,height=16cm]{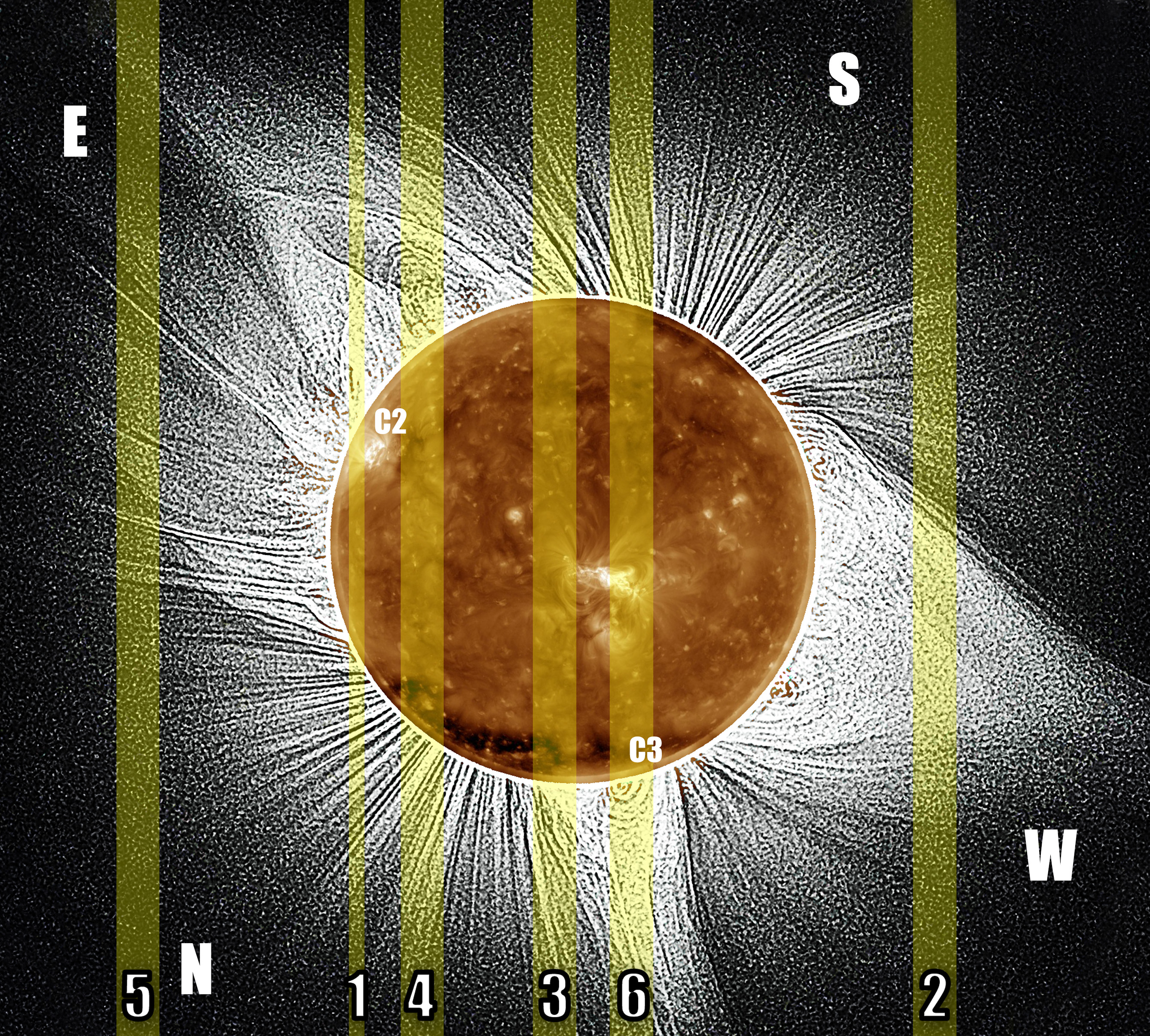}
    \caption{W-L coronal structures obtained from highly processed images by J. Mouette, combined with AIA/SDO 19.4 nm simultaneous image inserted with correct scaling. Yellow vertical bands show the regions covered by the different positions of the entrance slit during  totality, including the scanning transverse to the slit that corresponds to the diurnal motion during the burst of 10 spectra taken at each successive position 2, 3, 4, 5, and 6. Position 1 was used to obtain fast spectra (0.1 s exposure time) in the region of the C2 contact, see Fig. 1. The orientations correspond to a reversed image of the corona seen in the sky. The unprocessed image shows a large radial gradient with very bright inner parts, which mask the detail of the image shown on the right-hand side of Fig.\,\ref{fig:overview}.}
    \label{fig:composite}
\end{figure*}

 For the analysis of the corona with deep spectra, five sequences of ten consecutive spectra were launched at selected locations over the corona (Fig.\,\ref{fig:composite}). No guiding was made during each of the sequences, which were taken for each chosen typical location in the corona. Equatorial and polar regions were well covered by the long vertical slit. The uniform diurnal motion is used for the scanning in each selected part of the corona (see Fig.\,\ref{fig:composite}). The shift between two consecutive spectra of a sequence corresponds to approximately 10\,arcsec in the normal direction of the slit. Two sequences were taken far out in the middle corona. For all of these deep spectra, the exposure time was 1\,s and the duration of a sequence of 10 spectra was typically 14\,s. The resulting total amplitude of the scan produced by the diurnal motion is 125\,arcsec. Over the image of the corona, the width of the entrance slit corresponds to 54\,arcsec without taking the motion of the image on the slit into account during the exposure. The equivalent pixel size corresponds to 11\,arcsec, which gives, assuming a perfect focus, a resolution of 22\,arcsec along the slit. The demosaicing slightly affects the resolution along the slit by an amount smaller than the smearing due to the diurnal motion. Some overlapping then exists along the slit in the collected spectra for each selected region.

\begin{figure*}
    \centering
    \includegraphics[width=16cm]{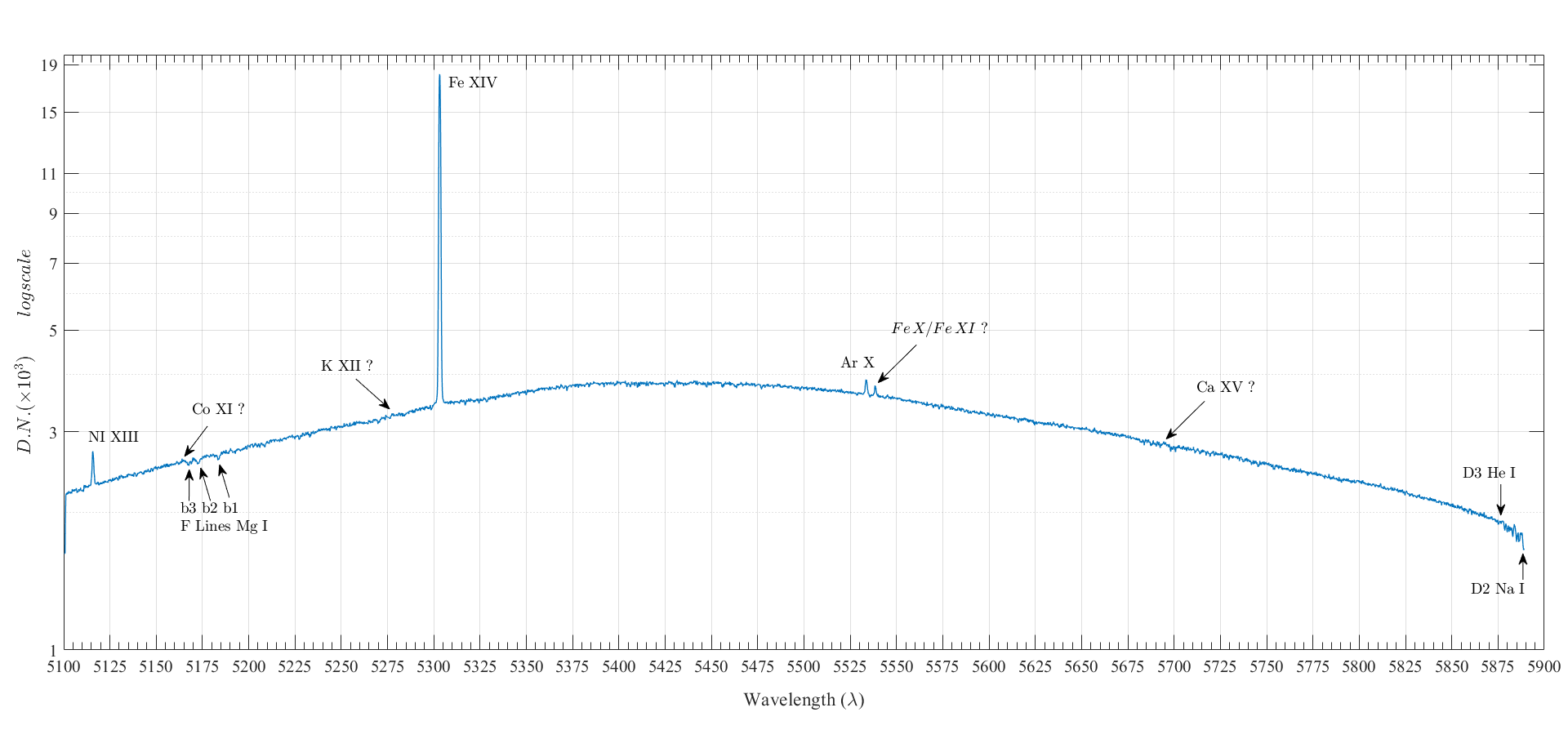}
    \caption{Summed spectra made at position 4 (see Fig.\,\ref{fig:composite}) that uses brightest part close to east equatorial region of corona. It shows the variation resulting from the spectral transmittance of the system and the typical behavior of the resulting spectra. It includes the signatures of the lines of the F-corona superposed on the K-corona continuum. The dark noise level and the offset are removed as well as the tiny distortion with respect to the chip. The whole spectrum corresponds to 5616\,px along the  wavelength axis.}
    \label{fig:specpos4}
\end{figure*}

The spectral sensitivity of the camera that we used relies on elementary pixels assembled in a matrix with color filters (also called the Bayer matrix). The effective spectral transmission of the spectrograph is mainly affected by first, the spectral sensitivity of green pixels that contribute the most (2\,px of the 4\,px Bayer matrix are covered with a green broad-band filter) because the central region of each spectrum is in the green region where the main coronal line of Fe\,XIV is located; second, the behavior of the solar spectrum with a maximum intensity in the blue-green region; third, the Earth's transmission is slightly better in the red part; and fourth, the spectral efficiency of the blaze of the holographic grating that we used, which is in the blue-green (500\,nm) region. The spectral region covers [510-590]\,nm where several interesting lines were expected, in addition to the most intense Fe\,XIV line that should be used for a deep analysis in the outer corona \citep{Allen46,Shklovskii65,Stellmacher74,Raju93,Contesse04,Koutchmy05}. Fig.\,\ref{fig:specpos4} shows the sum of raw spectra taken in position 4 of Fig.\,\ref{fig:composite}, using the bright part near the east equatorial region. It illustrates the variation in the summed spectral transmittance and intensity variations along the entire analyzed region. The behavior is typical of the spectral variations in our spectra of the inner corona, including the coronal emission lines and the signatures of F-lines of the F-corona superposed on the dominating continuum spectrum produced by the scattering of the solar disk intensities by the free electrons of the K-corona. It is the first time that the F-corona component has been recorded in this spectral region that includes the well-known Mg\,I triplet near 517.5\,nm that could produce a measurable depression in the K-corona continuum spectrum, which depends on the electron temperature \citep[e.g.,][]{Reginald14}.

\section{Results}

\subsection{The K and F spectra}
\label{sec:K+Fspec}

We collected 62 coronal 2D spectra covering several solar radii along the slit, including the line profiles of several emission lines. Our analysis concentrates on the most typical regions of both the inner corona and the middle corona, up to radial distances r = 2R$_{\odot}$ from the solar center and even further out. Since we are dealing with a corona close to the minimum of solar activity, two main parts around the Sun are prominent: first, equatorial regions with big loops and arches of the extended streamers, and second, the polar regions with coronal holes \citep[see Fig.\,\ref{fig:composite} and ][dealing with the modeling of coronal structures]{Mikic18}. Intermediate regions can also be discussed, such as the eastern active region and the edges of streamers or edges of CHs. Some dispersion results from the large integration effect along the l.o.s. \citep[e.g.,][]{Sornette80} that justifies this classification of the most typical parts of the corona.

\begin{figure*}
    \centering
    \includegraphics[height=7cm,width=16cm]{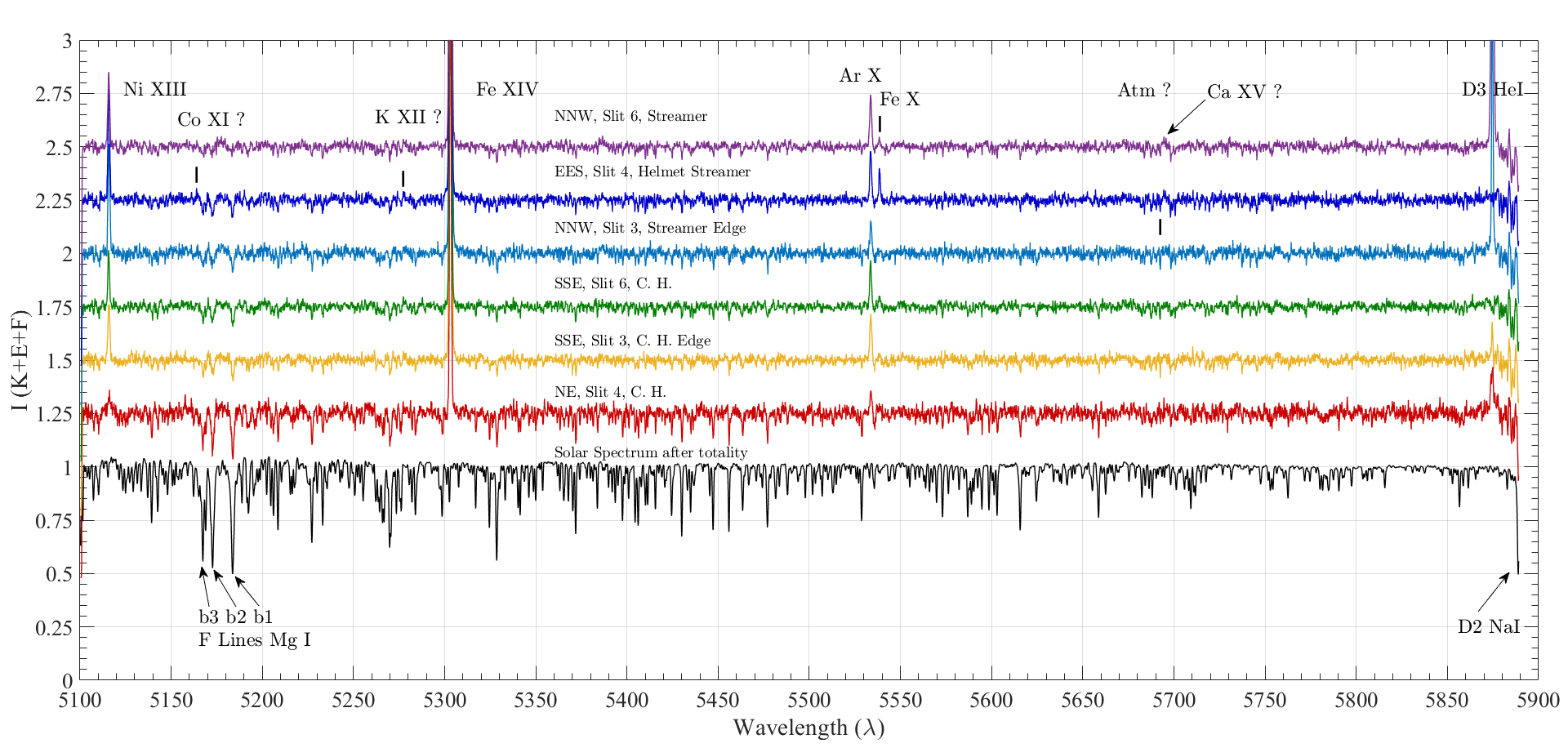}
    \caption{Full averaged K, F, and E corona spectra (on linear scale) for different regions after removing large scale spectral variations. A constant shift was introduced to make the comparison possible among the different regions. On the right-hand side, the strong D3 line of He\,I was recorded at the end of totality (C3 contact, position 6 at NNW) due to the slit crossing a small prominence.}
    \label{fig:specKFE}
\end{figure*}
\begin{figure*}
    \centering
    \includegraphics[height=7cm,width=16cm]{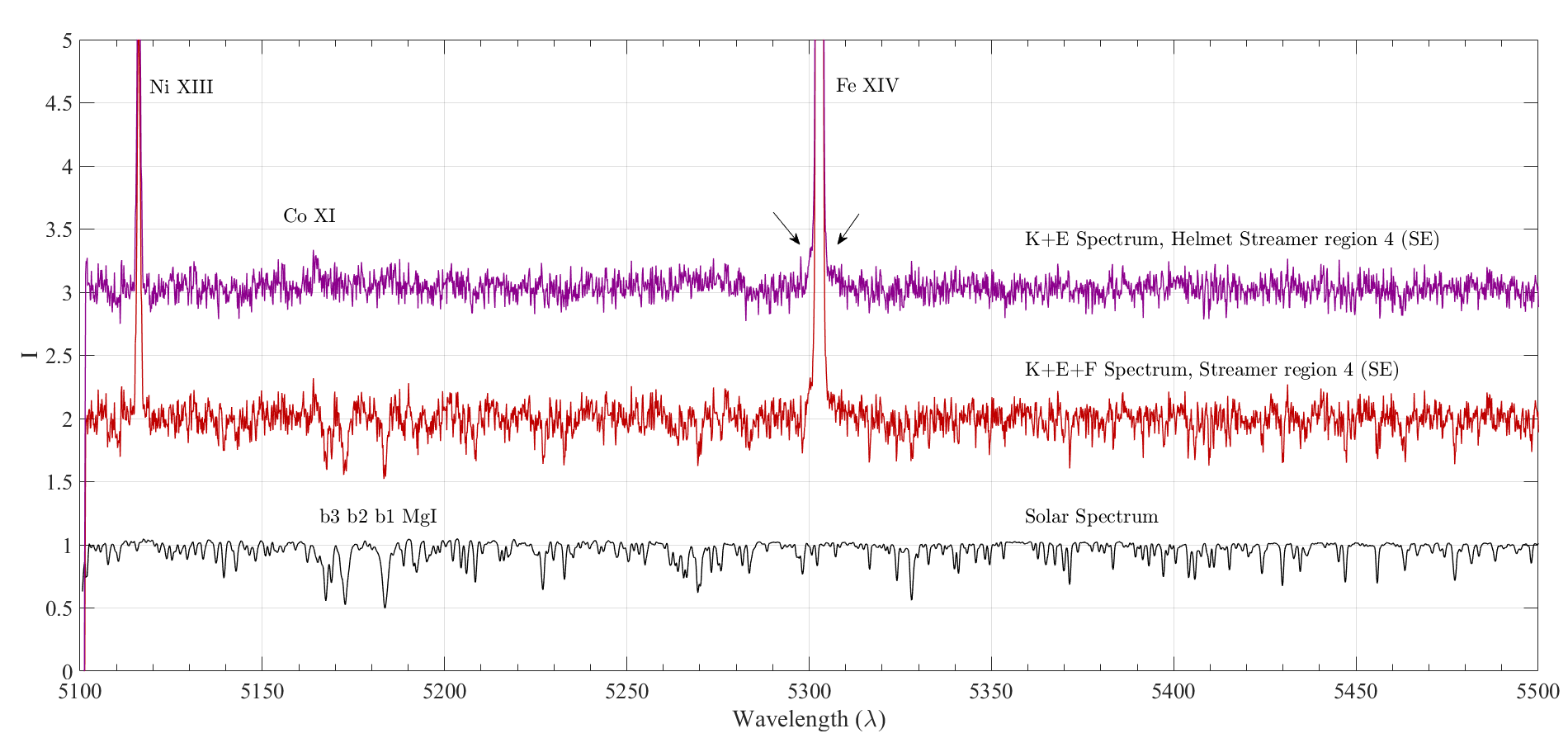}
    \caption{Selected part of spectrum of region 4 in east in green spectral region to illustrate procedure of removing F-corona component. The far wings of the Fe\,XIV line are seen near the center of the display.}
    \label{fig:specwings}
\end{figure*}

 The highest S/N is obtained when considering the brightest inner corona parts as seen in Fig.\,\ref{fig:overview}. No radial filter was used to get the spectra. The inevitable instrumental smearing in our spectra due to the use of a broad slit adds an effect in the behavior of the radial variation of intensities, making the inner parts similar to a sort of bright fringe in the spectra (see Fig.\,\ref{fig:A5}). The inner parts can be extended from the edge of the Moon to approximately one scale height in the corona (its radial extent is 70\,Mm or on the order of 100\,arcsec on average in the corona); nevertheless, it is important to bear in mind that the section that we consider in each spectrum is not taken with a slit in the radial direction (see Fig.\,\ref{fig:composite}). Finally, to produce the best possible spectra for each coronal region, we decided to sum the ten spectra taken at each position shown in Fig.\,\ref{fig:composite}; this induces some additional smearing that results from the scanning during observations, but it also increases the S/N. The analysis and discussion of the detail of each spectral sequence is left for a future work dealing with the modeling of individual structures. In Fig.\,\ref{fig:specKFE} we show the average spectra that cover the whole observed range, which is similar to Fig.\,\ref{fig:specpos4}, but we corrected for the variable gain along the wavelengths by filtering the low frequencies. Each region shown in Fig.\,\ref{fig:composite} where the slit crosses the edge of the Moon is considered. Spectra taken in the far corona (positions 2 and 5), with a much lower S/N, are not shown. In Fig.\,\ref{fig:specwings}, the blue part of the averaged spectrum of region 4 in the east (see Fig.\,\ref{fig:composite}) is shown to illustrate the method of removing the F-component from the corrected K, F, and E observed spectra. The F-corona is assumed to perfectly reflect the solar flux spectrum that we obtained a few minutes after the totality, which uses a diffuser positioned before the entrance aperture that does not change any optical parameters of the experiment. Different exposure times were used to optimize the S/N in the solar flux spectra. The amount of solar flux due to the F-corona in each spectrum is obtained by fitting the observed F-lines (see Fig.\,\ref{fig:specpos4} and \ref{fig:specwings}). The ratio F/K depends on the radial distance and the coronal regions because the intensity of the K-corona changes with the latitude. This ratio is near 6\% in the case that is illustrated in Fig.\,\ref{fig:specwings}. In the Appendix, Fig.\,\ref{fig:A1}, which is similar to Fig.\,\ref{fig:specwings}, shows spectra summed all around the disk. The F-component is clearly identified.\\
 Intensities given in Fig.\,\ref{fig:specKFE} and \ref{fig:specwings} are in relative units. It is of great interest to translate intensities in absolute units of the average disk brightness to compare the data with calculated line emissions from the emission measure analysis. A quick determination of intensities can be done from the 6\% F/K ratio evaluated in the case of the inner region 4 in the east (see Fig.\,\ref{fig:overview}), near r = $(1.15\pm0.05)$\,R$_{\odot}$ where a helmet streamer rises above. The determination is based on the use of the model of the F-corona of \citet{Koutchmy85}, assuming that the F-corona is constant in time on the scale of a solar cycle. We can readily deduce the absolute intensities of both the W-L continuum and of the emission lines shown in Table\,\ref{tab:table}.
 In position\,4 of Fig.\,\ref{fig:overview}, in the east, where F/K\,=\,0.06, we consider some average value in both the radial direction and in azimuth because the ten spectra are shifted by the diurnal motion.
 We obtained $B(K) = 1.5\,10^{-6} B_{\odot}$ where $B(K)$ is the K-corona brightness in units of  $B_{\odot}$ and the mean solar disk brightness \citep[which corresponds to $ \log{B(K)} = 4.18$ in units of $10^{-10}\,B_{\odot}$ that are usually adopted when comparing the radial behavior of the coronal brightness as in][]{Allen63}. This is close to the value given for the intensity of a minimum activity corona in equatorial regions. Intensities of the lines can then be evaluated, see Table 1, from the ratio $I_0$/(K+F) (see Table\,\ref{tab:table}). However, the absolute values are only useful in the framework of a comparison with computed values of the emission measure integrated along the l.o.s. making some simplifying assumptions, which is beyond the present analysis.

\subsection{New coronal lines}

\begin{figure*}
    \centering
    \includegraphics[width=16cm]{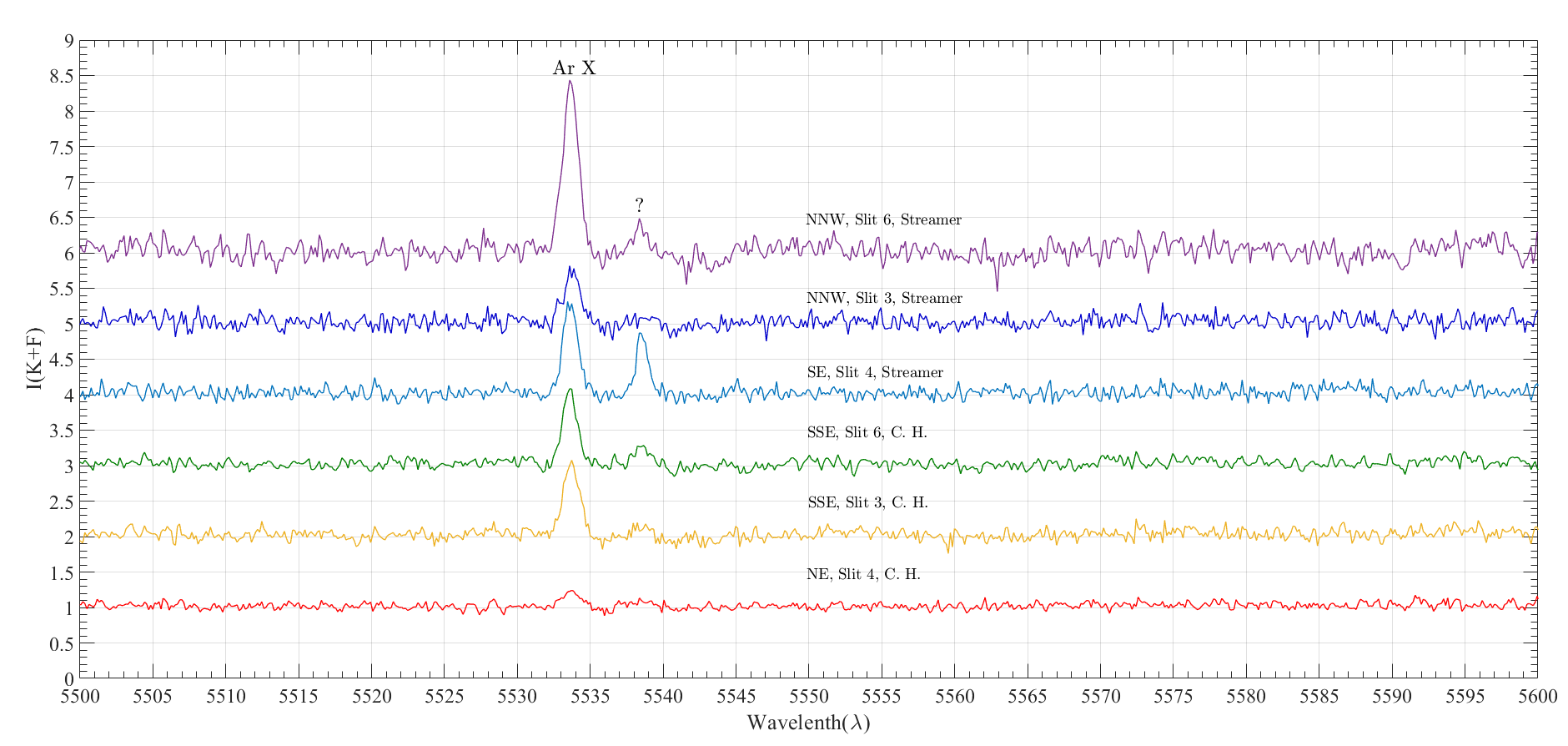}
    \caption{Selected part of our summed spectra in inner corona at different positions around Sun (see Fig.\,\ref{fig:composite}) to show intensity variation of new lines. Intensity is shown with a linear scale, with spectra shifted in the ordinate.}
    \label{fig:newlines}
\end{figure*}

A typical equatorial region spectrum mixes, along the l.o.s., the contributions from the active corona, with temperature and densities increased, and from a quiet corona made of intermittent small scale structures inside a low density corona. We also consider CH spectra, including contributions from polar plumes and jets of puzzling origin.\\ 
Some ambiguity results when a comparison is made with the predictions that modelers compute to obtain emission measures in the corona; this occurs because they usually do not mix regions of different temperatures and different densities. It makes the computed relative intensities of lines difficult to compare with the observed fluxes integrated along the l.o.s. However, the computations of the wavelengths of forbidden coronal lines \citep[e.g.,][]{Edlen43,Edlen69,Mason77} are a robust method of identifying the observed lines. We used the table of \citet{DelZanna18} to guide our identifications. In Fig.\,\ref{fig:newlines} and Fig.\,\ref{fig:A3}, we show a part of our spectra where new lines are definitely recorded, which are indicated by question marks in the table of \citet{DelZanna18}. The very interesting Ar\,X line at 553.35\,nm of this high FIP element is seen well in almost all parts of the inner corona, although it is not mentioned in the popular tables of coronal emission lines of many textbooks \citep{Allen63,Shklovskii65,Phillips92,Golub10} and poorly identified in other cases \citep[e.g.,][]{Mouradian97}. The line was already theoretically predicted in the seminal paper of \citet{Edlen43}, but it was not reported clearly in the literature \citep{DelZanna18} and its exact observed wavelength was not known. Thanks to our high S/N and our calibration procedure with the solar flux spectra (see Fig.\,\ref{fig:A3} and \ref{fig:A4}), we confirm its wavelength in air at 553.35\,nm (see Table\,\ref{tab:table}). This line was also noticed in TSE 2017 processed data from slitless spectrograph by A. Voulgaris (personal communication).

\begin{table*}[]
\centering
\def\arraystretch{1.5}
\begin{tabular}{|c|c|c|c|c|c|c|c|c|c|c|}
\hline
Wavelength & Ion & FWHM  & Ionization & FIP & $\mu$ & $\log$(A) & <$T_i$> & I$_0$/(K+F) & $T_{\rm eff}$ & V$_{nth}$ \\
 (\AA) &  & (\AA) & Potential (eV) & (eV) &  &  & (MK) &  & (MK) & (km/s) \\
\hline
5302.86 & Fe~XIV & 0.82$\pm$0.02 & 361 & 7.9 & 55.85 & 7.6 & 1.8 & 4.3 & 2.6$\pm$0.1 & 15.6$\pm$1.2 \\
5116.16 & Ni~XIII & 0.85$\pm$0.06 & 352 & 7.6 & 58.69 & 6.3 & 1.8 & 0.2 & 3.2$\pm$0.4 & 19.7$\pm$3.1 \\
5533.50 $\pm$0.10 & Ar~X & 1.03$\pm$0.12 & 422 & 15.7 & 39.9 & 6.8 & 2.0 & 0.08 & 2.9$\pm$0.7 & 17.2$\pm$6.7 \\
5538.30 $\pm$0.10 & Fe~X & 1.03$\pm$0.18 & 233 & 7.9 & 55.85 & 7.6 & 1.1 & 0.04 & 3.8$\pm$1.5 & ? \\
5164 & Co~XI & ? & 275 & 7.8 & 58.94 & 5.1 & ? & 0.01 & ? & ? \\
\hline
\end{tabular}
\vspace{2mm}
\caption{Parameters of coronal lines measured using summed spectrum taken in near equatorial region of  east limb, at radial distance of 3$'$. Wavelengths are given in air; the FWHM are corrected for the instrumental smearing; $\mu$ is the mass of the ion; A is the relative abundance of the ion; <$T_i$> is the temperature deduced from the ionization equilibrium; I$_0$ is the central intensity of the Gaussian profile of the line measured above the pseudo-continuum background of the K and F corona in units of the intensity of this background; $T_{\rm eff}$ is the temperature deduced from the FWHM without assuming any turbulent broadening; V$_{nth}$ is the non-thermal velocity (see Eq.\,\ref{eq:vnth}). The last row gives tentative values for the expected Co\,XI line recorded with an extremely low amplitude above the background level. The 5538.3\,\AA\, line is attributed to the ion Fe\,X, following a suggestion given in \citet{DelZanna18}.}
\label{tab:table}
\end{table*}

 Close to this prominent Ar\,X line (Fig.\,\ref{fig:specKFE}, \ref{fig:newlines}, \ref{fig:A2}, and \ref{fig:A3}), a fainter line is detected well in the southeast in position 4 and confirmed by a definite signature in other positions.  Its wavelength in air is 553.83\,nm. The line was barely suggested from different photographic observations \citep{DelZanna18}, although it does not appear in the extended table of \citet{Mouradian97} and it was not predicted in the seminal paper by \citet{Edlen43}. The line appears unidentified in the compilation given by \citet{Swensson74} in their Table\,IV with a wavelength at 553.91\,nm. From an observational point of view, it is clear that the earlier observations made with limited S/N were confused by the occurrence of two suggested lines at close wavelengths. Later in a paper devoted to a detailed theoretical study on the Ar\,X line, \citet{Edlen69} predicted in their Table\,4 that a coronal line at 553.9\,nm was possibly produced by the Fe\,XI ion. However \citet{Smitt77}, Table 2, in his revised theoretical study attributed this line to Fe\,X. It was again given in the revised Table\,V of \citet{Edlen78}. \citet{Mason77} suggested in their Table\,4 a weak line at 553.9\,nm, which was attributed to Fe\,XI. In our Table\,\ref{tab:table}, we provide a wavelength of 553.83\,nm for this line and conservatively attribute it to Fe\,X \citep{DelZanna18}. However, a line of Fe\,XI that corresponds to a significantly higher temperature of formation could be equally considered.  We observed this line best in the SE region of position 4,  corresponding to a small active region in the east limb (Fig.\,\ref{fig:composite}). It is a region where the line emissions of higher temperature would clearly be better observed.\\
Furthermore, we examine the part of the spectra where some very weak lines were episodically reported in the past. We do not analyze the very hot Ca\,XV at 544.4\,nm, which we barely observed just because it is a typical line of an active region where flares are observed, inside regions previously called a coronal condensation \citep[see][]{Aly62}. Our minimum activity corona does not show any flaring region at the limb. We also selected a line at 516.4\,nm that was never observed before \citep{Swensson74,Mouradian97,DelZanna18} but was repeatedly mentioned in the literature since 1942 \citep{Edlen43} as being possibly produced by Co\,XI, a low FIP element. The line is seen just above the noise level in Fig.\,\ref{fig:specwings} in the K and E spectrum. The line appears only after the removal of the contribution of the F-corona and it is not seen in the CH regions. We give some parameters for this extremely weak line in our Table\,\ref{tab:table}.

\subsection{Analysis of the width of the coronal lines and the turbulent velocities}

\begin{figure*}
    \centering
    \includegraphics[width=16cm]{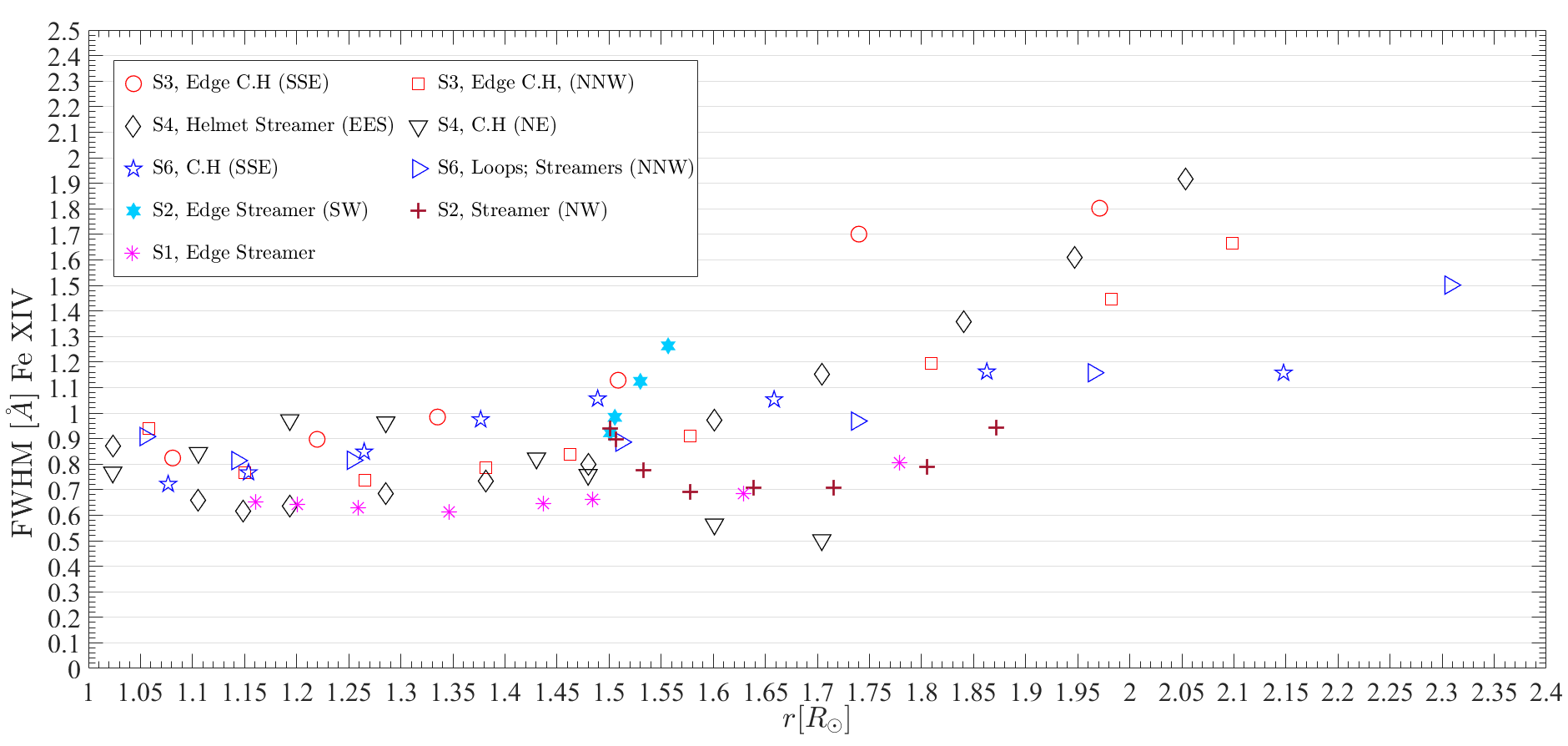}
    \caption{Variation with radial distances in measured FWHM corrected for instrumental broadening using averaged line profiles of each sequence for all positions of slit as shown in Fig.\,\ref{fig:composite}.}
    \label{fig:width}
\end{figure*}

\begin{figure*}
    \centering
    \includegraphics[width=16cm]{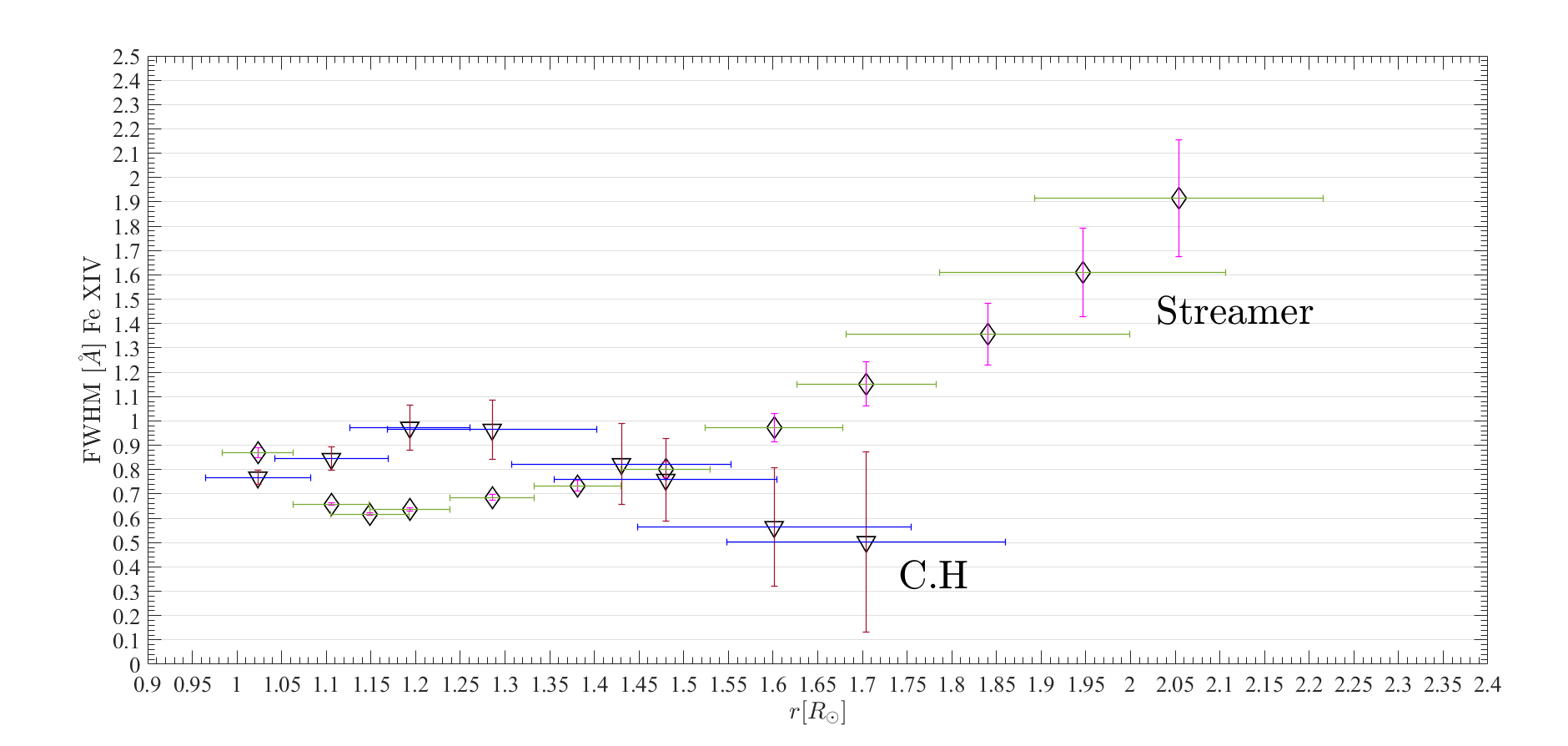}
    \caption{Radial variation of corrected FWHM in position 4 for streamer region and for CH region. The slit crosses the upper part of the streamer region close to the boundary of the south CH.}
    \label{fig:width2}
\end{figure*}

The intensity profiles $I(\lambda)$ of each line were accurately fit with a Gaussian function to determine its full width at half maximum\footnote{The so-called Doppler width is sometimes used with a value FWHM/1.665 \citep[e.g.,][]{Wilhelm11}.} (FWHM, noted $\Gamma$ hereafter) after subtracting the local background (see Fig.\,\ref{fig:A4}). The measured FWHM ($\Gamma_m$) were then used to determine the so-called non-thermal width \citep[$\Gamma_{nth}$, see, e.g., ][]{Tsubaki75,Doyle98,Kim00,Koutchmy05}. The instrumental FWHM ($\Gamma_{inst}$) was accurately evaluated from the solar flux spectra, which was made immediately after the totality by comparing it with a much better resolved solar flux spectrum from the Sacramento Peak Observatory. The smearing in our spectra is deduced as a Gaussian function with a FWHM $\Gamma_{inst}=0.072$\,nm. Correcting our profiles by removing the instrumental broadening permits the estimate of the so-called effective temperature T$_{\rm eff}$ \citep{Kim00}, which is also called the Doppler temperature \citep{Tsubaki75}, see Table\,\ref{tab:table}.\\
To compute the non-thermal velocities $V_{nth}$ corresponding to our profiles, we needed the thermal broadening ($\Gamma_{th}$) of the corresponding lines. The ion temperature $T_i$  is usually extracted from the ionization equilibrium curves, giving the most probable temperature of the ion \citep{Phillips92}. Table\,\ref{tab:table} shows the values we assumed. The $V_{nth}$ are then given by:
\begin{equation}
V_{nth} = \sqrt{\frac{1}{4\ln 2} (\frac{c}{\lambda})^2 (\Gamma_m^2 -\Gamma_{inst}^2) - V_{th}^2 }
\label{eq:vnth}
\end{equation}
with $c$ the speed of light and $\lambda$ the wavelength of the line. Thermal velocities are given by:
\begin{equation}
V_{th} = \sqrt{\frac{2kT_i}{\mu m_p}}
\end{equation}
with $\mu$ the relative mass of ions (see Table\,\ref{tab:table}), $T_i$ their temperature, and $m_p$ the mass of a proton.\\
The computed $V_{nth}$ for the inner corona in the equatorial region observed in position 4 (see Fig.\,\ref{fig:composite}) is given in Table\,\ref{tab:table}. The table also contains $T_{\rm eff}$ , which is of possible importance because it includes the non-thermal (turbulent) broadening.\\
Fig.\,\ref{fig:width} shows the FWHM of the Fe\,XIV line for all observed regions around the solar disk and further out into the corona at larger radial distances, which were corrected for instrumental broadening. Fig.\,\ref{fig:A5} illustrates the method of extracting the values at different radial distances. Intensities and S/N drastically decrease with radial distances as illustrated in Fig.\,\ref{fig:A6}. In Fig.\,\ref{fig:width2}, we show the behavior of the FWHM for the position 4 only in order to compare a streamer region and a CH region. The values for the CH region are, however, noisier for distances larger than 0.5\,R$_{\odot}$ from the limb. 


 \section{Discussion and perspectives}
 \label{sec:conc}

Although we concentrated our analysis on the coronal emission lines, we first notice the high S/N of our original spectra. In Fig.\,\ref{fig:specpos4} showing the whole spectrum, which is uncorrected for the spectral transmittance intensity variation but corrected from the dark current level and the offset, the level of the continuum corresponds to typically 3.10$^3$ counts. This allowed us to determine the ratio of intensity of the F-corona to the intensity of the K-corona close to the limb. The  dust component near the bright inner edge of the corona, at a radial distance on the order of 1 arcmin to 3\,arcmin from the Sun, has been controversial in the past \citep{Koutchmy73} partly due to the suggestion of a dust free zone around the Sun with an unknown extent. The forthcoming measurements that the Parker Solar Probe will soon collect when flying in the neighboring area of the Sun should bring some in situ evaluation that hopefully will shed some light on this question.\\
Accordingly, it is possible to separate the intensities due to the K and F components thanks to their very different spectral signatures. The method of matching the superposed F spectrum seen over the continuum spectrum of the K-corona that uses the solar flux spectrum, which is measured immediately after the totality with a high S/N, gives excellent precision \citep[e.g.,][]{Ichimoto96}. This method originally permitted the discovery of this F-component thanks to the signature of the broad H and K lines of Ca\,II that are imprinted over the near-UV eclipse photographic spectra \citep{Allen46}; however, it was never used in the green part of the coronal spectrum. A more complete analysis taking all observed positions around the Moon into account will allow us to deduce a more relevant value of the F/K ratio when using spectra of 1\,s exposure time.\ This leads to a better resolution instead of averaged values as given in Section\,\ref{sec:K+Fspec}. We also plan to use our absolute calibration from well-known bright stars superposed on our white-light images, which were obtained simultaneously to analyze intensity variations in more detail.\ Nevertheless, that is outside the scope of this paper.

The more immediate result of this study is the discovery and the definite confirmation of several coronal emission lines. The analysis of the high FIP and rather hot line of Ar\,X with an ionization potential that is significantly higher than the green line of Fe\,XIV (see Table\,\ref{tab:table}) is of great interest when discussing the abundance ratio in the active parts of the corona. Its intensity is, not surprisingly, lower inside the CH regions (see Fig.\,\ref{fig:specKFE}). Its non-thermal velocity amplitude does not seem to be different from the $V_{nth}$ of other well observed low FIP lines (Fe\,XIV and Ni\,XIII, see Table\,\ref{tab:table}). Our method of averaging ten spectra that come from adjacent regions of the corona (streamers base, equatorial region, CHs, polar regions, and intermediate regions) may hide some effect. We plan to carry out a detailed analysis for each sequence to look at the line parameters for single regions before interpreting intensities, and we will compare them to the emission measures simultaneously observed in EUV filtergrams obtained with the AIA of the SDO mission.

As shown by Fig.\,\ref{fig:specKFE}, the Fe\,XIV line is observed everywhere in the corona, even in the darkest regions of the deep polar CHs. This is clearly a result of the improved S/N compared to previous studies and especially compared to the Lyot coronagraph analysis \citep[e.g.,][]{Contesse04} and even studies made in space \citep{Mierla08}. A part of the Fe\,XIV flux is seen in CH regions that come from distant regions, which contribute along the l.o.s. However, another part could come from hot plumes during their active phase, thus producing extended polar jets in W-L and in coronal lines \citep[e.g.,][]{Wang98,Tavabi18}. We plan to search for this signature again by using AIA filtergrams.

The non-thermal velocities inside the inner corona reported in Table\,\ref{tab:table} show that they are not significantly different for different lines (we exclude the last Fe\,X/Fe\,XI line at 553.83\,nm with a very low intensity). Amplitudes are dispersed around a value on the order of $\pm$\,17\,km/s. This value is in agreement with the classical values reported from high resolution photographic spectra \citep[e.g.,][]{Nikolsky71} and the values reported by \citet{Kim00}, which were based on Fabry-Perot measurements. The value is marginally smaller than the values coming from Lyot coronagraphic spectroscopic measurements \citep[e.g.,][]{Contesse04} and the values reported from space-borne instruments \citep[e.g.,][]{Doyle98,Gupta17}. The non-thermal velocities are often discussed in the context of heating-related coronal waves \citep[e.g.,][]{Wilhelm11,Brooks16}. A possible diagnostic of the wave heating mechanism could be the behavior of waves with a radial distance from the surface since they are presumably magnetic waves with a decreasing value of the magnetic field and of plasma densities with height. In Fig.\,\ref{fig:width}, which shows the radial behavior of the Fe\,XIV linewidths, an increase is seen above R\,=\,1.5 to 1.6\,R$_{\odot}$, corresponding to a significant increase of the waves' amplitude. The effect is confirmed by using a more detailed analysis (see Fig.\,\ref{fig:width2}), although the measurements inside the CH above 1.5\,R$_{\odot}$ become difficult. Another possible interpretation of the FWHM becoming larger for R\,>\,1.6\,R$_{\odot}$ is the integrated contributions along the l.o.s. of small blobs or plasmoids flying away toward the solar wind \citep[e.g.,][]{Sheeley97,Jones09,Tavabi18}. This distance near 1.6\,R$_{\odot}$ (or 0.6\,R$_{\odot}$ above the solar surface) corresponds to the top of the loops that are at the feet of the streamers from where small clouds of plasma slowly fly away as described by \citet{Tavabi18}. We are then dealing with quasi-radial flows giving a small velocity component along the l.o.s., which could contribute to the so-called slow solar wind.

In our case the averaging procedure is important when discussing Doppler shifts, as well as the large width of the slit, which mixes the phases of the waves along the l.o.s. However, thanks to the S/N in the Fe\,XIV green line observations, we can look at the net Doppler shifts of the full line using the center of gravity method (position 3 southeast of Fig.\,\ref{fig:composite}, see also Fig. A7). A drift toward the blue, corresponding to a relative velocity up to 4\,km/s, is observed close to the limb (streamer), and the shift seems to be reversed for R\,>\,1.4\,R$_{\odot}$ (south CH). We preliminarily interpret the shifts to be produced by a combination of the flows and of the rigid rotation of the magnetic corona.

Another aspect that is relevant to coronal physics comes from the inspection of the feet of the Fe\,XIV line in the inner corona: the non-Gaussian component of the profiles (Fig.\,\ref{fig:specwings} and \ref{fig:A4}). The interpretation of these extended wings of the line, which are made visible by the summing of the profiles taken at several positions and along the l.o.s., is not straightforward. We suggest effects of the contribution of unresolved features seen far enough from the plane of the sky with significant proper motions as the most likely explanation. The width of these feet corresponds to velocities on the order of $\pm$\,50\,km/s and more, their proper motion could be even larger due to the projection effect if the motion is nearly radial as this could be the case in the vicinity of CHs. The corona that we observed is close to a minimum type corona and we can expect permanent contributions of polar CHs. A more detailed analysis, and more observations, should be carried out along the radial direction to see if the non-Gaussianity increases together with the FWHM of the profiles, but unfortunately the S/N will also decrease drastically with the radial distance.

 To conclude, we note that there is the need for better spatio-temporal sequences of line profiles with a similar spectral resolution as our present observations, at least in the inner corona, in order to cover the analysis of smaller features including loops and polar plumes. We plan to look at the Fe\,XIV line profiles in the outer corona where the mechanism of excitation of the line becomes collisionless and photo-excitation dominates \citep{Shklovskii65,Allen75,Koutchmy05,Habbal11} which is where important new physical mechanisms could explain the outward flows at the origin of the solar wind. A comparison will be provided with the imaging data including the large amount of material collected in space during this eclipse. Finally, our analysis deals with a quasi-minimum corona where CHs are observed well but both the plasma densities and the coronal temperatures are significantly lower than during periods when the Sun is active. The total eclipse in Dec 2020 and even more so, those of 2024 and after, will open the way to study the active solar corona. In space, a giant coronagraph using a formation flying mission with a spectroscopic capability \citep[as originally proposed for the Proba\,3 mission of ESA, ][]{Lamy08} would gather essential information due to the lack of coronal emission line profile analysis in the R\,>\,1.5\,R$_{\odot}$ corona. This is needed to expand upon the results that were collected with the LASCO\,C1 coronagraph of the SoHO mission \citep{Mierla08}.

\begin{acknowledgements}
AIA images are from the SDO mission (NASA). We would like to warmly thank Xavier Jubier for providing the lunar profile and ephemerides computed for our site. Jean Mouette (IAP) supplied the high resolutions W-L images of the eclipse corona taken from the same site. John Stefan (NJIT) provided inner corona pictures from his polarization analysis. We thankfully acknowledge G. Del Zanna for providing some enlightening comments. We also benefited from discussions with B. Filippov, E. Tavabi, J.-L. Leroy, J.-C. Vial, C. Bazin, and from the review by L.M.R. Lock and the anonymous referee. S.K. was supported by the Harvard-Smithsonian Institution through his participation to the AIR-spec project. Some support was also provided by the French LATMOS thanks to a PNST contract. The setup at the McDonald's ranch (Indian Valley, Idaho) benefited from the help of the ranch owners. 
\end{acknowledgements}

\bibliographystyle{aa}
\bibliography{ecl17}

\begin{appendix}

\section{Appendix}
Complementary spectra and images are given to better illustrate the data processing and to show the phenomena with more detail and in a different way. 

\begin{figure*}
    \centering
    \includegraphics[height=8cm,width=12cm]{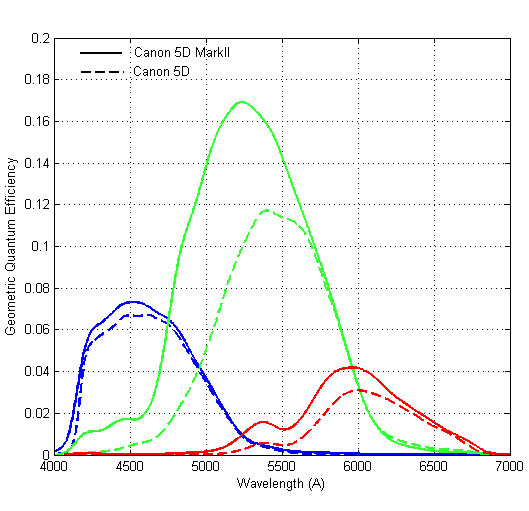}
    \caption{Spectral sensitivity of camera used for these observations (C. Buil, private communication, see \texttt{http://www.astrosurf.com/buil/50d/test.htm}).}
    \label{fig:sensi}
\end{figure*}

\begin{figure*}
    \centering
    \includegraphics[height=7cm,width=16cm]{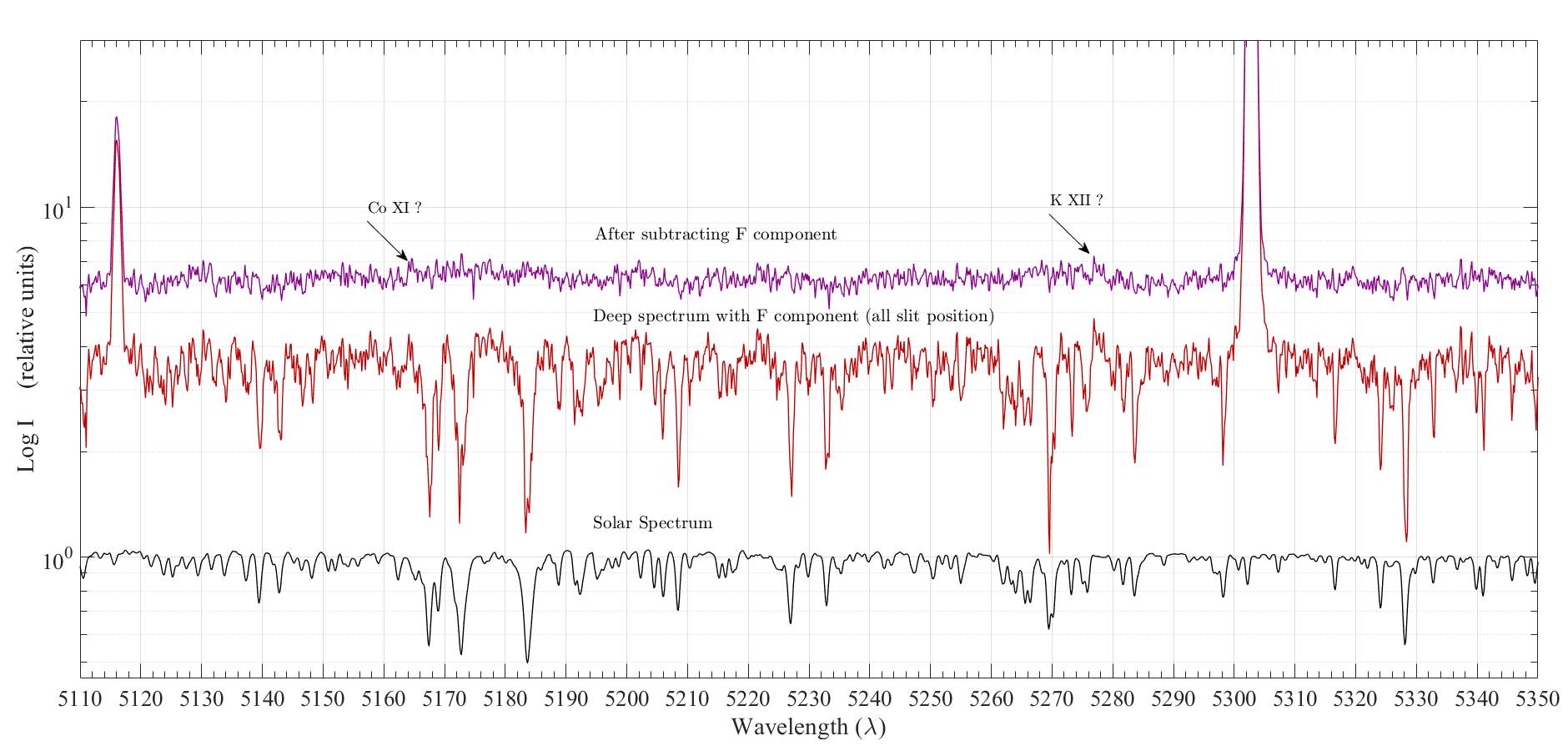}
    \caption{Spectra similar to Fig.\,\ref{fig:specwings}, but computed by summing all contributions from inner corona around disk (see Fig.\,\ref{fig:composite}) and using log scale. It includes the contributions from the CH regions where the relative contribution of the F-corona is much higher.}
    \label{fig:A1}
\end{figure*}

\begin{figure*}
    \centering
    \includegraphics[height=7cm,width=16cm]{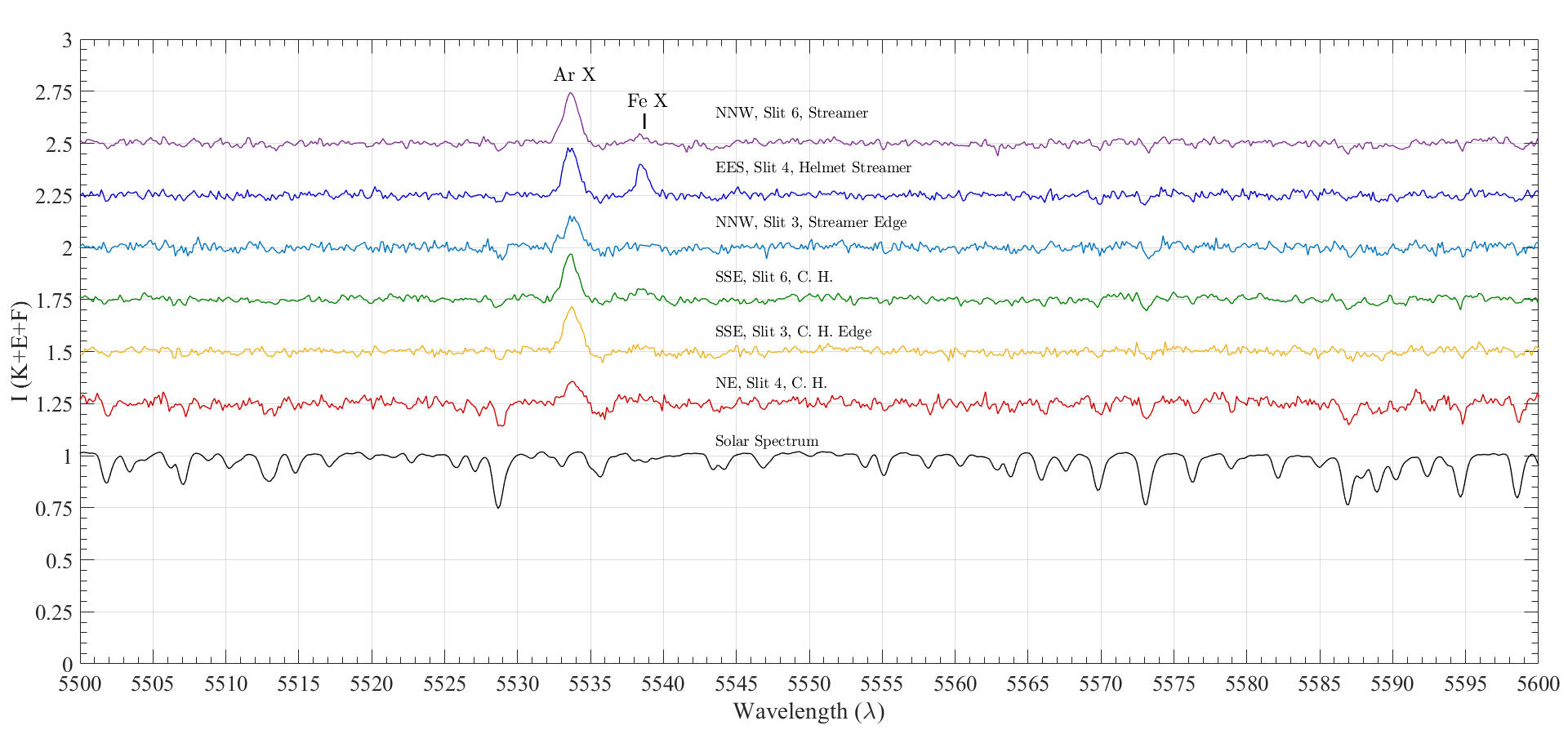}
    \caption{Spectra similar to Fig.\,\ref{fig:newlines} in linear scale and without removing F-component, with calibration solar flux spectrum shown at bottom. }
    \label{fig:A2}
\end{figure*}

\begin{figure*}
    \centering
    \includegraphics[height=6cm,width=16cm]{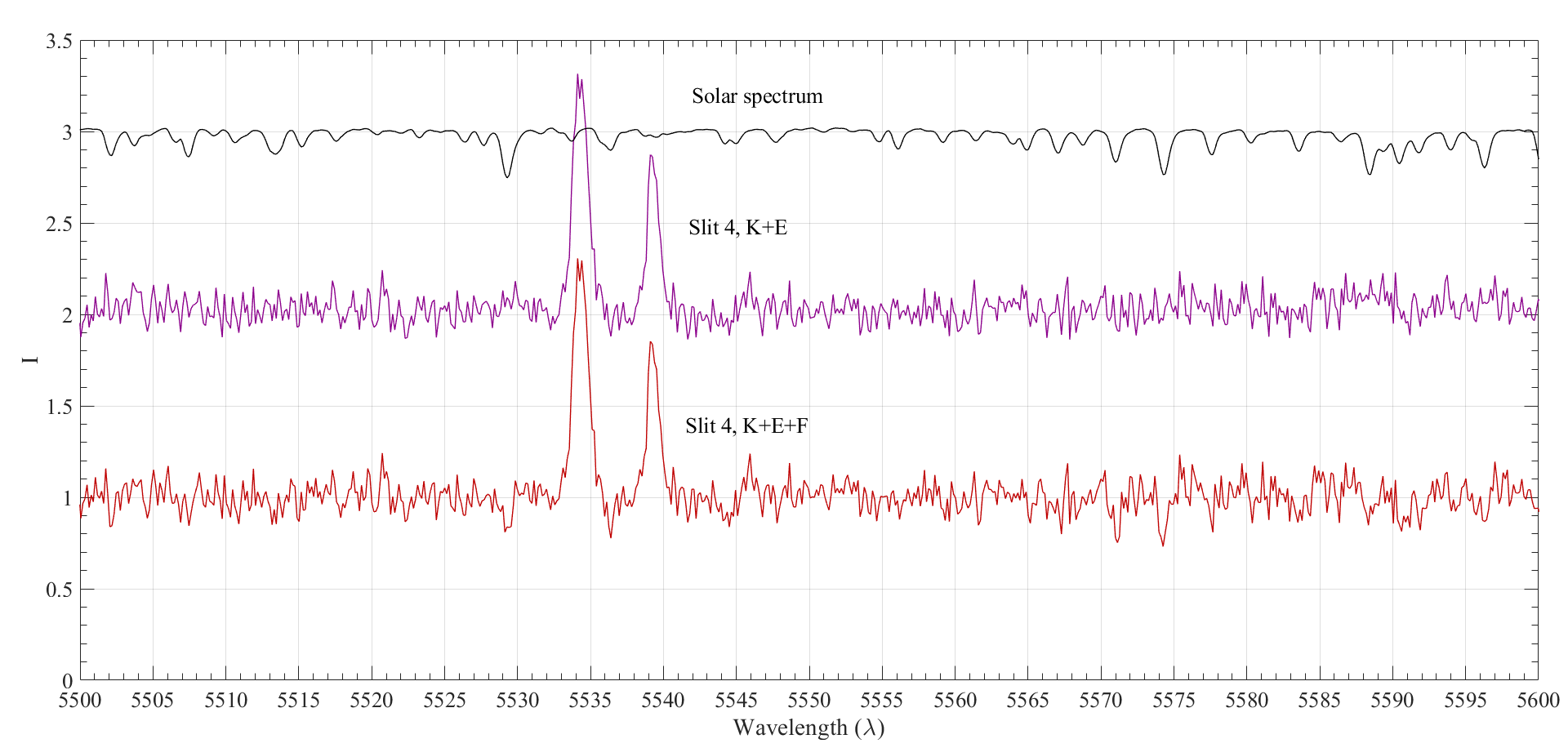}
    \caption{Partial K, F, and E spectra with Ar\,X line compared to solar flux spectrum to illustrate goodness of wavelength calibration. The spectra are shifted by a small amount to best match the solar spectra obtained after the totality. Some telluric lines on the right-hand side make the comparison difficult.}
    \label{fig:A3}
\end{figure*}

\begin{figure}
    \centering
    \includegraphics[height=7cm,width=9cm]{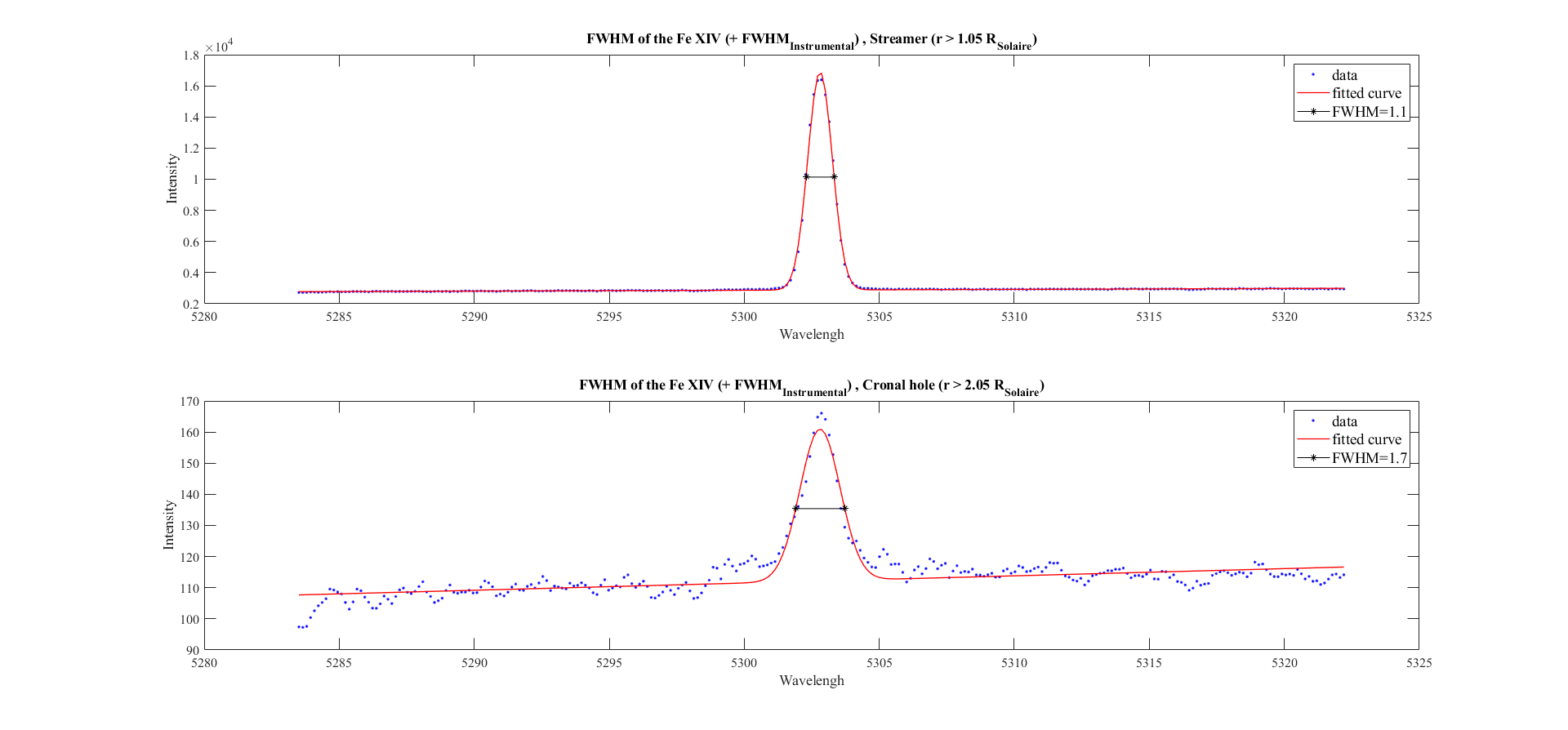}
    \caption{Examples of fitting intensity profiles Fe\,XIV line observed at different location inside corona extracted from sequence 4 of Fig.\,\ref{fig:composite}, upper part in southeast. The top profile (uncorrected for the slit broadening width is 0.11\,nm) is taken in the inner corona where the slit crosses the big loops and arches at the foot of a big streamer; the bottom profile (width is now 0.18\,nm) is taken at more than 1.05\,R$_{\odot}$ from the limb (R\,>\,2.05\,R$_{\odot}$) and corresponds to the slit crossing the edge parts of the streamer near the boundary of the polar CH in the south. Some wing non-Gaussianity could be seen in the bottom profile.}
    \label{fig:A4}
\end{figure}

\begin{figure}
    \centering
    \includegraphics[height=10cm,width=9cm]{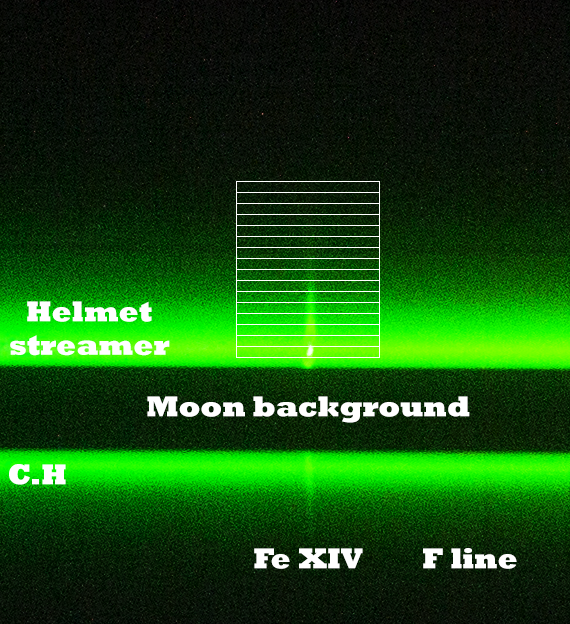}
    \caption{Image of part of single spectrum near green Fe\,XIV line to schematically illustrate how successive measurements of FWHM are made inside helmet streamer in position 4 of Fig.\,\ref{fig:overview}. The bottom part corresponds to the coronal hole in the north.}
    \label{fig:A5}
\end{figure}

\begin{figure*}
    \centering
    \includegraphics[height=8cm,width=15cm]{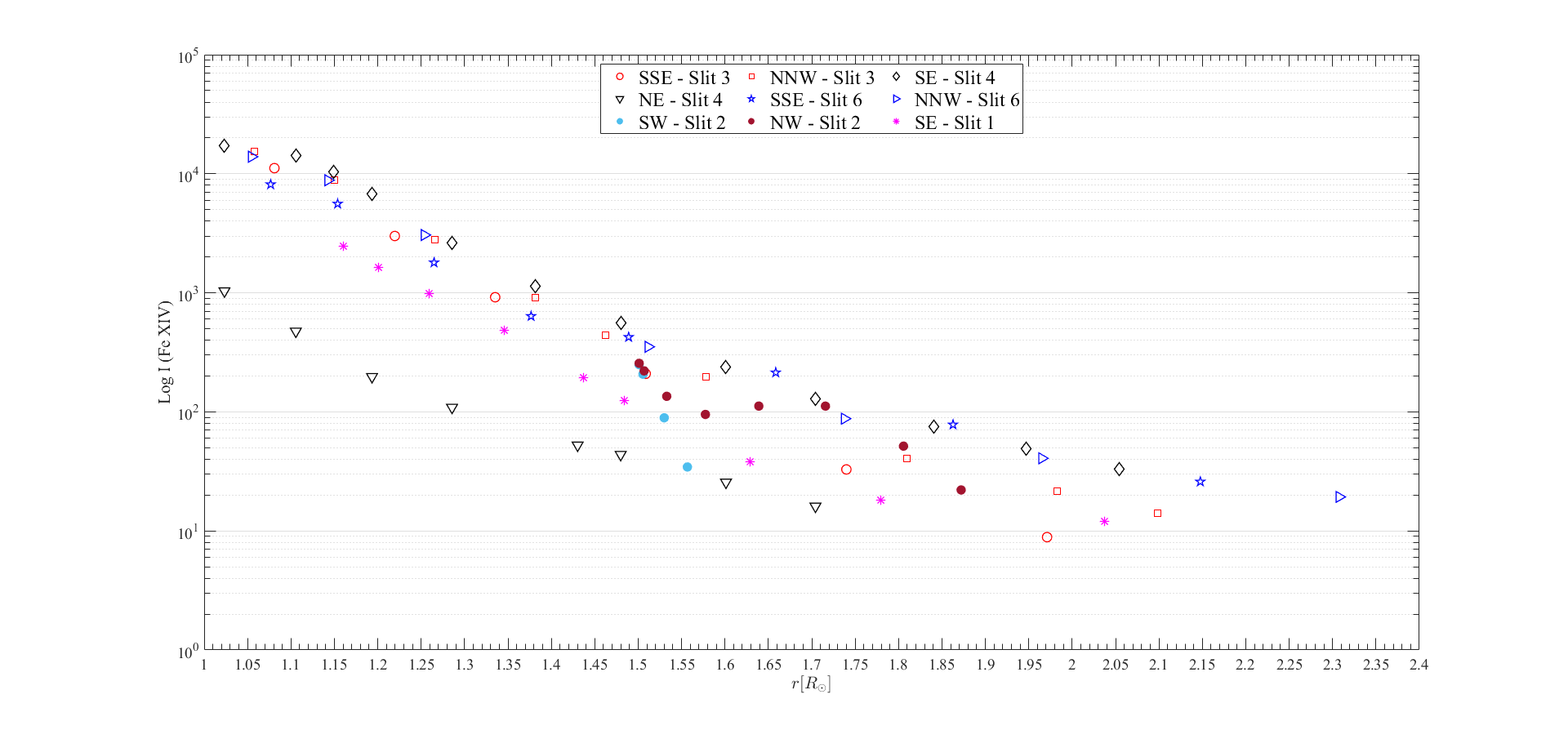}
    \caption{Intensity variation of  Fe\,XIV line for different positions in corona and radial distances. The log scale covers three decades of intensity variations measured above the local K and F background. The line is measured well everywhere, including the deep CH region of position 4 in Fig.\,\ref{fig:composite}.}
    \label{fig:A6}
\end{figure*}

\begin{figure}
    \centering
    \includegraphics[height=6cm,width=9cm]{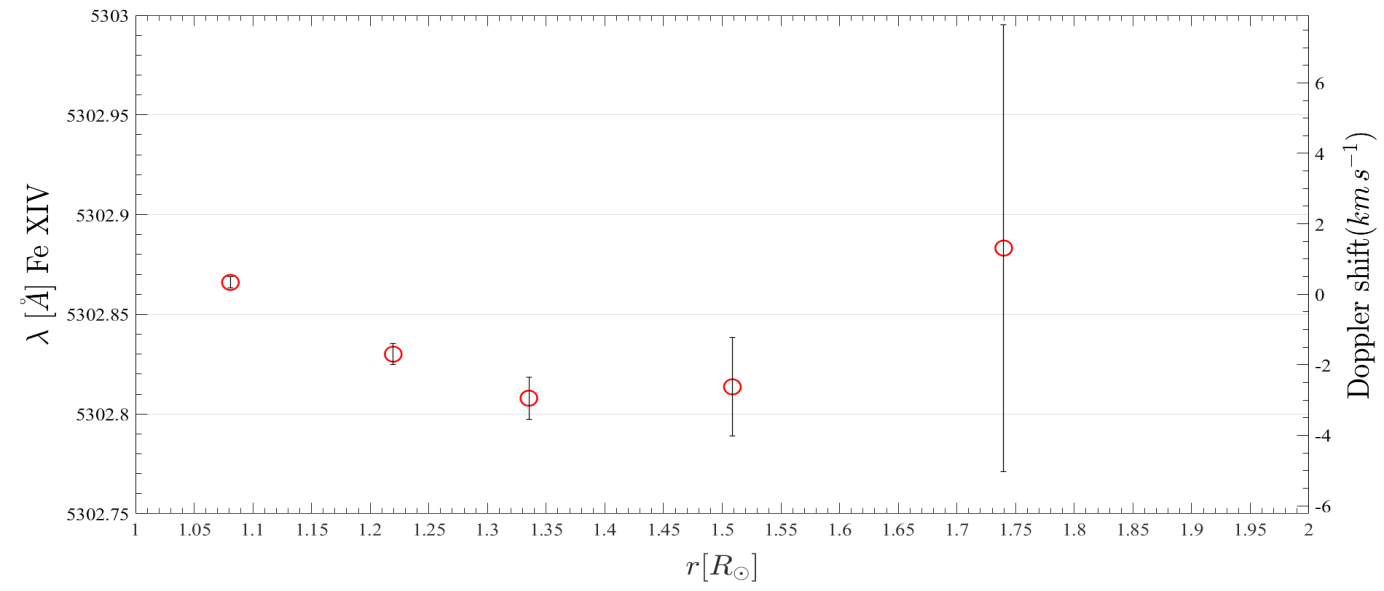}
    \caption{Central wavelength of Fe\,XIV line observed along radial direction (position 3 of Fig.\,\ref{fig:composite}, up to SE) using method of center of gravity of full line profile. Values for the very inner corona do not show an error bar at this scale. On the right-hand side, the corresponding scale for Doppler velocities are shown.}
    \label{fig:A7}
\end{figure}

\end{appendix}

\end{document}